\magnification=1200
\hsize=150.truemm
\vsize=220.truemm
\baselineskip=2.0em
\parskip=4.truemm
\parindent=5.truemm
\rm

\vskip1.0truecm

\centerline{  \bf   The $L$-Matrix for the Massive Thirring Model }
\vskip1.0truecm

\centerline{ Y. K. Zhou$^{1,2,3}$\footnote{}
    {\sevenrm 3.Postal Address after 1 April,1992:}
\footnote{}{\sevenrm Math. Dept, Melbourne Univ, Parkville,Vic 3052, Australia}
            and K. D. Schotte$^1$ }

\vskip .25in
\centerline{ $^1$FB-Physik, FU-Berlin, Arnimallee 14, 1000 Berlin 33, FRG}  
\vskip .10in 
\centerline{ $^2$Institute of Modern Physics, Northwest University} 
\centerline{ Xian 710069, P. R. China}
\vskip .10in
\vskip3.0truecm

\centerline{ \bf Abstract }
   
   As the new results for the massive Thirring model the $L$-matrix and the
algebraic relations for its action angle variables are given.
So it is shown most directly that this model which describes self-interacting
relativistic Fermions in one-dimensional space is a quantum integrable system.
\vskip6.truemm
\line{PACS numbers: 1110, 0290, 0365, 0370. \hfill }

\vfill
\null
\eject

\vskip1.0truecm
\line{ \bf 1 Introduction \hfill }

  In this note we present the $L$-matrix for the massive Thirring model [1]
defined by the Hamiltonian
$$H=\int_0^M dx \{\phi^+(-i\sigma_3)\partial_x\phi +m\phi^+ \sigma_1 \phi
   -2\sin (2c)\phi^+_1 \phi^+_2 \phi_2 \phi_1 \} , \eqno(1)$$
where $\sigma_i$ are the Pauli matrices. The Fermion fields $\phi (x)=\bigl(
\phi_1(x),\phi_2(x)\bigr)$ satisfy the usual equal time anti-commutation 
relation
$\{\phi_{\mu}(x),\phi^+_{\nu}(y)\}=\delta_{\mu \nu}\delta (x-y)$, the indices 
refer to right or left moving particles. The constant $c\le {\pi \over 2}$
gives the strength of interaction between the Fermions. As a solvable and 
simple model the massive Thirring model has attracted much interest (see refs.
[2-11]).

   The $L$-matrix corresponding originally to the Lax-pair for nonlinear partial
differential equations is an operator that determines a "canonical" 
transformation of the field variables $(\phi^+,\phi)$ to variables of 
action-angle type, which are the matrix elements of the so-called monodromy
matrix. The integrability can be shown by constructing a $R$-matrix, which 
gives the information about the algebraic relations between these action-angle
variables. Hence the $L$-matrix and the $R$-matrix are crucial for the study 
of an integrable system.

  The Thirring model is equivalent to the quantum sine-Gordon model [2]. For
this bosonic model the $L$-matrix and $R$-matrix have been given by Sklyanin,
Takhtadzhyan and Faddeev [4]. We think it is an interesting problem to 
construct these matrices also for its fermionic counterpart especially since
supermatrices are the tools which are not so common with integrable systems.

  Here we will study the $L$- and $R$-matrices for the Thirring model. The
method we used originally is to find the $L$-matrix by taking the continuum
limit from the $R$-matrix of an inhomogeneous six vertex model. It is
known that the six vertex model can be used to construct the Bethe ansatz
solution of the massive Thirring model [8] and that the Hamiltonian should be 
connected to a six vertex model with staggered weights [10]. Actually, we 
used this method before to derive the $L$-matrices for the sine-Gordon model 
[13] and a bosonic system with the hamiltonian similar to equ. (1) [12]. Since 
in the present case the derivation is lengthy and similar to the one 
in [12], we use here a different approach. Given the result, we have to prove
that we have found the $L$-matrix of the Thirring model. So we have to find 
the $R$-matrix starting from the $L$-matrix 
in order to show the integrability. Since $L$-matrix determines the monodromy 
matrix, we have to check whether the latter contains really the Hamiltonian 
given above as one of the simple conserved entities.
\vskip1.0truecm

\line{\bf 2. $L$-matrix \hfill }
    
    The action-angle variables or more specific the monodromy matrix $T(u)=
T(x=M|u)$, $M$ is the length of the system, are defined by the differential
equation
$$ {dT(x|u)\over dx}=:L(x|u)T(x|u): \eqno(2)$$
and the boundary condition $T(x=0|u)=I$, the identity matrix. The colons mean
normal ordering of the Fermi operators and the $u$ is the spectral parameter.

  We found for the $L$-matrix after taking the continuum limit and introducing
fermion fields
 $$L(x|u)=i{1\over 2}m\sinh u\tau_3 +\Sigma (x) +S(x|u) . \eqno(3)$$
Here we distinguish the Pauli matrices in classical space $\tau_i $ and
those in quantum space  $\sigma_i $ used before. For $\Sigma $ and $S$ we have
 $$\Sigma (x)=
    \left( \matrix{ -\phi^+(x)(1-e^{ic\sigma_3})\phi(x), & 0 \cr
    \noalign{\vskip2pt}
             0, &-\phi^+(x)(1+e^{-ic\sigma_3}) \phi(x)  \cr}\right) ,\eqno(4)$$

 $$S(x|u)=i\sqrt{m\sin c}\left(\matrix{
          0, & e^{-{u\over 2}}\phi_1(x)-e^{u\over 2}\phi_2(x)  \cr
-e^{-{u\over 2}}\phi_1^+(x)+e^{u\over 2}\phi_2^+(x), &0 \cr} \right). \eqno(5)$$
We have to choose the $L$-matrix as a super-matrix with its row or column 
parities $p(1)=0$ and $p(2)=1$, consequently $T(x|u)$ as a solution of equ.(2)
is also a super-matrix.

  The integrability of the Thirring model can be shown by finding the 
Yang-Baxter equation for the monodromy matrix $T(u)$
 $$R(u-v)T(u){\buildrel \otimes \over {\sevenrm s}} 
         T(v)=T(v){\buildrel \otimes \over {\sevenrm s}}T(u) R(u-v) ,\eqno(6)$$
where the tensor product indicated by the symbol ${\buildrel \otimes \over 
{\sevenrm s}}$
is of the super-form $(A{\buildrel \otimes \over {\sevenrm s}}B)^{k,l}_{i,j}
\break
=(-1)^{p(j)(p(k)+p(i))} A_{i,k}B_{j,l}$. 
The matrix $R$ has the same form as for the six
vertex model, note however that the spectral parameter $u$ is imaginary in
the usual six vertex model
 $$ R(u)=\left(\matrix{
\sinh({u\over 2}+ic), & 0                 &0,              &0             \cr
             0,        &\sinh ic ,     &\sinh {u\over 2},    &0             \cr
             0,     &\sinh {u\over 2},     &\sinh ic ,       &0             \cr
             0,         &0,                  &0,     &\sinh({u\over 2}+ic)\cr}
                                  \right) . \eqno(7)$$

   In the following let us prove the Yang-Baxter equation (6). We can rewrite
the differential equation (2) as an integral equation for $0\le x\le M$
$$\eqalignno{ T(x|u)=&e^{i{m\over 2}x\sinh u\tau_3}   & \cr
 &+\int_0^x dz e^{i{m\over 2}(x-z)\sinh u\tau_3}
  :(\Sigma (z)+S(z|u))T(z|u): , &(8) \cr}$$
from which we 
can get more easily two auxiliary equations 
$$\eqalign{ &\phi_{\nu}(x)T(x|u)=\tau_3 T(x|u)\tau_3 \phi_{\nu}(x)+
   {1\over 2}:E_{\nu}(x|u)T(x|u):   , \cr
           &T(x|u)\phi^+_{\nu}(x)=\phi^+_{\nu}(x)\tau_3 T(x|u)\tau_3 +
          {1\over 2}:F_{\nu}(x|u)\tau_3 T(x|u)\tau_3 : ,\cr }  \eqno(9)$$
with
$$\eqalign{
    &E_{\nu}(x|u)=-\phi_{\nu}(x)(1-\tau_3 e^{ic\tau_3 (-1)^{\nu +1}})+
     i\sqrt{m\sin c}(-1)^{\nu}e^{(-1)^{\nu}{u\over 2}}\tau^- \cr
    &F_{\nu}(x|u)=-\phi^+_{\nu}(x)(1-\tau_3 e^{ic\tau_3 (-1)^{\nu +1}})-
    i\sqrt{m\sin c}(-1)^{\nu}e^{(-1)^{\nu}{u\over 2}}\tau^+  .\cr }\eqno(10)$$

  Defining the tensor product in the Yang-Baxter equation (6) as
       $$K(x|u,v)=T(x|u){\buildrel \otimes \over {\sevenrm s}}T(x|v), 
                          \eqno(11) $$
one can get using the last equs. (9) and (10) a differential equation for this
tensor product
   $$\partial_x K(x|u,v)=:L (x|u,v) K(x|u,v):   \eqno(12)$$
with a $L$-matrix depending now two spectral parameters $u$ and $v$ 
  $$L(x|u,v)=L(x|u){\buildrel \otimes \over {\sevenrm s}}1+
             1{\buildrel \otimes \over {\sevenrm s}}L(x|v)+
 \sum_{\nu } F_{\nu}(x|u){\buildrel \otimes \over {\sevenrm s}}E_{\nu}(x|v).
                         \eqno(13)  $$
One can show only by an explicit calculation that
 $$R(u-v) L(x|u,v)=L(x|v,u) R(u-v) . \eqno(14)$$
The equation means that $R$-matrix can exchange $u$ and $v$ in $L(x|u,v)$ in
manner the Yang-Baxter relation postulates. The equs. (11-13) above defining 
the tensor product $K(x|u,v)$ from $L(x|u,v)$ are of course also valid if the 
spectral parameter $u$ and $v$ are exchanged. This implies that an equation 
like equ. (14) must hold also for $K$
 $$R(u-v) K(x|u,v)=K(x|v,u)R(u-v) , \eqno(15) $$
which is the Yang-Baxter relation (6) taking $x=M$.
 
  Hence the transfer matrix $t(u)=T(u)_{11}-T(u)_{22}$ which is the super-trace
of the $T(u)$ must commute for different spectral parameters, i.e.
   $$[t(u),\phantom{-} t(v)]=0 . \eqno(16)$$
So the definition of the $L$-matrix (3) generates a quantum integrable system.
The main problem is now to show that this quantum system is the massive 
Thirring model.

\line{\bf 3. Hamiltonian and momentum \hfill }
 
  Here we will find the Hamiltonian of the Thirring model (1) from the transfer 
matrix $t(u)$. This shows directly that the $L$-matrix (3) gives the Thirring
model.
Using an integral form of equ. (2) similar to (8), however expanding $T$ with
respect to $S$ given by equ. (5), one has
   $${\tilde T}(x|u)=:Q:+\int_0^x dze^{-i m z\tau_3 \sinh u}
                                      :Q(z)S(z|u){\tilde T}(z|u): ,
             \eqno(17)$$
where ${\tilde T}(x|u)=e^{-i{m\over 2}x \tau_3 \sinh u}T(x|u)$, 
$Q(z)=\exp (\int_z^x dy \Sigma (y))$ and $Q=Q(z=0)$.
In order that such an expansion makes sense one must add
an imaginary part to the spectral parameter, for example, 
$u\to u\pm i{\pi\over 2}$ to have simple expressions. Thus from
  $$\eqalign{  T(x|u\pm i{\pi\over 2})=
&e^{i{m\over 2} x \tau_3 \sinh (u\pm i{\pi\over 2})}:Q: +\cdots \cr
 &=e^{\mp{m\over 2}x \tau_3 \cosh u}:Q: +\cdots  \cr }\eqno(18)$$
we can see that the transfer matrix $t(u\pm i{\pi\over 2})$ decrease or
increase rapidly for $u \to \pm \infty $. After iterating equ. (17) 
we see that the expansions 
 $$\eqalign{  {\tilde T}(x|u-i{\pi\over 2})_{11}=:Q_{11}: 
  +m\sin c\int_0^x dz_1 &\int_0^{z_1}
      dz_2e^{-m(z_1-z_2)\cosh u } \cr   
  &\times : \Phi_- (z_1)\Phi^+_- (z_2) Q_{11}: + \cdots \cr}  \eqno(19a)$$
and
 $$\eqalign{  {\tilde T}(x|u+i{\pi\over 2})_{22}=:Q_{22}:  
          +m\sin c\int_0^x dz_1& \int_0^{z_1}
       dz_2e^{-m(z_1-z_2)\cosh u }  \cr 
  &\times : \Phi^+_+ (z_1)\Phi_+ (z_2) Q_{22}: + \cdots \cr} \eqno(19b)$$
have nontrivial limits for $u\to \pm \infty $, where 
$$\Phi_{\pm } (z)=Q(z)_{11}\bigl( e^{-{u\over 2}\mp i{\pi \over 4}}\phi_1(z)-
           e^{{u\over 2}\pm i{\pi \over 4}}\phi_2(z)
                    \bigr) Q^{-1}(z)_{22} $$ 
and
$$\Phi^+_{\pm } (z)=Q(z)_{22}\bigl( e^{-{u\over 2}\mp i{\pi \over 4}}\phi^+_1(z)
  -e^{{u\over 2}\pm i{\pi \over 4}}\phi^+_2(z)\bigr) Q^{-1}(z)_{11} . $$ 
By choosing the sign of the imaginary part of $u$ we have picked out the 
exponential growing contributions. Neglecting 
the decreasing part related to $\exp (-e^{\pm u})$ for $u \to \pm \infty $,
the transfer matrix multiplied by
$\exp \{\pm i{m\over 2}M\sinh (u\pm i{\pi\over 2})\}$ is simply 
${\tilde T}(u-i{\pi \over 2})_{11}$ or $-{\tilde T}(u+i{\pi \over 2})_{22}$ in
this limiting case. We combine the contributions defining the generators
 $$\eqalignno{
  G_{\pm}(u)=\lim_{u\to \pm \infty } &\{e^{\pm ic}\ln 
       [e^{-i{m\over 2}M\sinh (u-i{\pi\over 2})}t(u-i{\pi \over 2}) & \cr
  &-e^{\mp ic} \ln (e^{i{m\over 2}M\sinh (u+i{\pi\over 2})}
        t(u+i{\pi \over 2})]\}. &  (20)\cr}$$
The generators $G_{\pm }$ can be calculated from equs. (19) using partial 
integrations repeatedly. In this way one obtains a series in $e^{\mp u}$ for 
$u\to \pm \infty $
  $$G_{\pm}(u)=\sum_{s\ge 0} C_{\pm s} e^{\mp su}   . \eqno(21)$$
The calculation is tedious but straightforward. Every term of the expansions
${\tilde T}_{11}$ and ${\tilde T}_{22}$  has 
contribution to $G_{\pm }$, even to the first order $C_{\pm 1}$. 
Fortunately, all coefficiences of the factor $e^{\pm u}$ can be summed up and 
give us the wanted results 

  $$C_{+1}={8\sin c\over m}\int_0^M dz[i\phi^+_2 \partial_z \phi_2
    +{m\over 2}(\phi^+_1 \phi_2 +\phi^+_2 \phi_1 )-\sin (2c)
     \phi^+_1 \phi^+_2 \phi_2 \phi_1 )] \eqno(22)$$ 
and
  $$C_{-1}={8\sin c\over m}\int_0^M dz[-i\phi^+_1 \partial_z \phi_1
    +{m\over 2}(\phi^+_1 \phi_2 +\phi^+_2 \phi_1 )-\sin (2c)
     \phi^+_1 \phi^+_2 \phi_2 \phi_1 )]. \eqno(23)$$ 
It can be seen that the generators $G_-$ and $G_+$ give the 
conserved quantities for right and left moving particles respectively. 
The Thirring system includes both contributions of right and left 
particles. The sum of the two first order coefficients is just
the Hamiltonian of the Thirring model (1) whereas the difference is the 
momentum
   $$P=-i\int_0^M dz \phi^+(z) \partial_z \phi (z) .   \eqno(24) $$
The other coefficients $C_{\pm s}$ $(s=0,2,3,\cdots )$, if they are not zero,
should give other conserved quantities of the Thirring model. These are not 
easy to calculate and cannot be studied here.
\vskip1.0truecm
\line{ \bf 4. Conclusion \hfill}
 
   In this note we have described the $L$- and $R$-matrices for the
Thirring model which were not given before for a fermionic relativistic theory.
For quantum inverse scattering transformations the $L$- 
and $R$-matrices are important operators, for example, for studying the 
algebraic Bethe ansatz and the inverse problem of the Thirring model in the 
sense of the works [4] and [14]. Also it is interesting to note that the   
Yang-Baxter equation (6) for the monodromy matrix has a super structure,
i.e. the diagonal elements of the monodromy matrix are of bosonic type, whereas
the off-diagonal ones are of fermionic type. Hence equation (6) gives us
also a graded and deformed algebra or a graded quantum algebra (see refs.
[15,16]).

\line{ \bf  Acknowledgement \hfill}

   One of the authors, YKZ, wishes to thank W. Nahm, E. Olmedilla, 
V. Rittenberg and Al. B. Zamolodchikov for discussion. He is also grateful 
to the Alexander von Humboldt Foundation for the award of a Research Fellowship.

\vfill
\null
\eject

\line{ \bf References  \hfill}
\item{1.} W. Thirring, Ann. Phys. 3(1958)91; F. A. Berezin and V. N. Sushko,
          Sov.--JETP 21(1965)865.
\item{2.} S. Coleman, Phys. Rev. D11(1975)2088.
\item{3.} A. Luther, Phys. Rev. B14(1976)2153.
\item{4.} E. Sklyanin, L. A. Takhtadzhyan and L. D. Faddeev, Teor. i. Mat.
          Fiz. 40 (1979)194.
\item{5.} V. E. Korepin, Teor. Mat. Fiz. 41(2)(1979)169.
\item{6.} H. Bergknoff and H. B. Thacker, Phys. Rev. D19(1979)3666.
\item{7.} H. B. Thacker, Rev. Mod. Phys. 53(1981)253.
\item{8.} M. F. Weiss and K. D. Schotte, Nucl. Phys. B225(1983)247.
\item{9.} T. T. Troung and K. D. Schotte, Nucl. Phys. B220(1983)77; B230(1984)1.
\item{10.}C. Destri and H. J. de Vega, Nucl. Phys. B290(1987)363; Mod. Phys.
          Lett. 4(1989)2595.
\item{11.}H. J. de Vega, Adv. Stud. in Pure Math. 19(1989)567 and the references
          therein.
\item{12.}Y. K. Zhou, Phys. Lett. B (1992) in press.
\item{13.}Y. K. Zhou, Int. J. Mod. Phys. A (1992) in press.
\item{14.}F. C. Pu and B. H. Zhao, Nucl. Phys. B275(1986)77.
\item{15.}P. P. Kulish and N. Yu Reshetikhin, Lett. Math. Phys. 18(1989)143.
\item{16.}H. Saleur, Saclay PhT/89-136.

\vfill
\eject
\end